# Profitable Strategy Design for Trades on Cryptocurrency Markets with Machine Learning Techniques

Mohsen Asgari[#1], Seyed Hossein Khasteh[#2]

[#]Artificial Intelligence Department, Faculty of Computer Engineering, K. N. Toosi University of Technology

[1]`Mohsen0Asgari@gmail.com`

[2]`khasteh@kntu.ac.ir`

*Abstract*— **AI and data driven solutions have been applied to different fields and achieved outperforming and promising results. In this research work we apply k-Nearest Neighbours, eXtreme Gradient Boosting and Random Forest classifiers for detecting the trend problem of three cryptocurrency markets. We use these classifiers to design a strategy to trade in those markets. Our input data in the experiments include price data with and without technical indicators in separate tests to see the effect of using them. Our test results on unseen data are very promising and show a great potential for this approach in helping investors with an expert system to exploit the market and gain profit. Our highest profit factor for an unseen 66 day span is 1.60. We also discuss limitations of these approaches and their potential impact on Efficient Market Hypothesis.**

## I. INTRODUCTION

Artificial Intelligence has widely been applied to different areas in the last decade and there have been reported a lot of improvements in results by using its applications. One of the very interesting areas of application is financial markets. There could be a lot of improvements in exploiting these markets by means of artificial intelligence and machine learning. Some cases of these applications include Loan Credit Scoring, Credit Evaluation, Sovereign Credit Ratings, Mortgages Choice Decision, Portfolio Management, Financial Performance Prediction and Market Direction Prediction (Bahrammirzaee, 2010).

In this paper we focus on the Market Direction Prediction problem and we look at it as a data science problem.

One of very innovative usages of new technology in finance is cryptocurrencies. As we read in (Hileman & Rauchs, 2017, p. 2) "The findings are both striking and thought-provoking. First, the user adoption of various cryptocurrencies has really taken off, with billions in market cap and millions of wallets estimated to have been 'active' in 2016. Second, the cryptocurrency industry is both globalised and localised, with borderless exchange operations, as well as geographically clustered mining activities. Third, the industry is becoming more fluid, as the lines between exchanges and wallets are increasingly 'blurred' and a multitude of cryptocurrencies, not just bitcoin, are now supported by a growing ecosystem, fulfilling an array of functions." which has been quoted from a survey by Cambridge Center for Alternative Finance in 2017, a year which is not even comparable to what is the prevalence of blockchain based technologies now. As of the time of writing this, only BTC makes a market cap of $1,042,689,199,152 and a 24 hour circulating volume of $57,794,818,577 (Today's Cryptocurrency Prices by Market Cap, 2021). As the dominance of BTC is 50 percent (Today's Cryptocurrency Prices by Market Cap, 2021) at the moment, the total market cap of cryptocurrencies registered in Coin Market Cap database can be estimated as about more than 2 thousand billion US Dollars. This impressive amount of money shows a great potential for this innovative use of technology with increasingly facing new challenges (like expensive transaction fees) and tackling them over with innovative solutions (like MicroCache (Almashaqbeh, Bishop, & Cappos, 2020)).

The main goal of the methods described in this article are to determine if the price of the analysed cryptocurrencies will move higher or lower in the coming four hours. To do that we use a data channel directed to the Binance cryptocurrency exchange free available API and receive the data from the exchange's database. Then we run some preprocessing procedures on them and get them ready to be used as entry to our machine learning models.

Three different machine learning methods have been used in this work. The first one is kNN as an example-base learner and the last two ones being Random Forest and Gradient Boosting methods as tree-base learners. These models have been discussed in the "Methods and Materials" section.

Data used for these analyses are Open, High, Low, Close and Volume data from three different cryptocurrency markets: ETH-USDT, LTC-BTC, ZEC-BTC. We have two sets of experiments for each model, one with this data being augmented by technical indicators and one without them.

After explaining the models and the data, we explore the implementation of these models in the "Proposed Method" section.

At the "Experimental Results" section we look at the performance of these models in predicting the next four hour movement of the market in the test data, which our learned models have not been exposed to.

At the "Discussion" section we look at the performance of our models and we discuss some different and debatable aspects of these methods and the whole strategy design system and their relation to Efficient Market Hypothesis. There are some improvements which can be made to this work and we mention some of them in the "Conclusion and Future Works" section.

## II. RELATED WORKS

In this section we introduce three different surveys done on the topic of market direction prediction and also point to the previous usages of the implemented methods in other studies.

First Survey (Bustos & Pomares-Quimbaya, 2020, p. 8) shows a comprehensive taxonomy of stock market prediction algorithms based on machine learning and their categories. This survey has also a performance comparison of the stock market forecast models (Bustos & Pomares-Quimbaya, 2020, p. 10) which has 47 different models compared with each other. Based on findings of this article the interest in using Market Information and Technical Indicators as inputs to models have increased in the past few years. It also shows more attention to ensemble methods for this topic recently. Another interesting finding in this survey is better accuracy obtained by using Technical Indicator and Social Networks data combined together in comparison with other data sources as input.

Second survey (Obthong, Tantisantiwong, Jeamwatthanachai, & Wills, 2020) points out advantages and disadvantages of using 23 different machine learning models with each other, which include k-NN and Random Forest. k-NN, described as a Classification and Forecasting Algorithm, has been noted to have advantages of being robust to noisy training data and being very efficient if the training datasets are large. It also points to the issue of determining the best k for this algorithm and its high complexity in computation and memory limitations as its disadvantages. k-NN can be sensitive to the local structure of the data based on the findings in this survey (Archana & Elangovan, 2014) (Jadhav & Channe, 2016). In the same survey, random forest has been categorized as another Classification and Forecasting algorithm and for its advantages we read: "Robust method for forecasting and classification problems since its design that is filled with various decision trees, and the feature space is modelled randomly, automatically handles missing values and works well with both discrete and continuous variables". RF algorithm has been disadvantaged by the following points "Requires more computational power and resources because it creates a lot of trees and requires more time to train than decision trees" (Obthong, Tantisantiwong, Jeamwatthanachai, & Wills, 2020, p. 5) (Pradeepkumar & Ravi, 2017).

Third survey (Kumar, Jain, & Singh, 2021) organises core Computational Intelligence approaches for stock market forecasting in three different classes including: Neural Network, Fuzzy Logic and Genetic Algorithm. It surveys application of these models in markets of 19 different countries. Mostly used data for training models based on this survey are Technical Indicators (Kumar, Jain, & Singh, 2021, p. 15). It also shows that more research has been done for American Markets (NYSE & NASDAQ) than other geographical locations. This survey concludes "identification of suitable pre-processing and feature selection techniques helps in improving the accuracy of stock market forecasting models and computational intelligence approaches can be effectively used to solve stock market forecasting problem with high accuracy. Among them hybrid models are predominant techniques applied to forecast stock market due to combined prediction capability of base models".

k-Nearest Neighbours algorithm (k-NN) is an instance-base learner model first developed by (Fix, 1985). This model has shown a good performance regarding returns in financial markets. Applying this model to Jordanian Stock Market has been reported to yield Total Squared RMS error of 0.263, RMS error of 0.0378 and the average error of -5.434E-09 for "AIEI" symbol (Alkhatib, Najadat, Hmeidi, & Shatnawi, 2013). Another scheme of applying k-NN has been reported in (Chen & Hao, 2017) by the name of "FWKNN". It has been concluded in that research: "The experiment results clearly show that FWSVM-FWKNN stock analysis algorithm where the classification by FWSVM and the prediction by FWKNN, is robust, presenting significant improvement and good prediction capability for Chinese stock market indices over other compared model".

Random Forests have been used since the late 90s to overcome the over fitting problem in decision trees (Ho, The random subspace method for constructing decision forests, 1998). A variation of this algorithm has been applied to cryptocurrency market direction detection problem on 60 minutes data in (Akyildirim, Goncu, & Sensoy, 2021). Their out-of-sample accuracy on BTC, ETH, LTC and ZEC has been reported 0.52, 0.53, 0.53 and 0.52 respectively. They have used mostly OHLC and indicator-based data for their model training. They also have concluded that their used algorithms "demonstrate the predictability of the upward or downward price moves" (Akyildirim, Goncu, & Sensoy, 2021, p. 27).

Gradient Boosting is a relatively old popular machine learning method in dealing with non-linear problems (Friedman, 2001). Later a more efficient variant of it has been developed by (Chen, et al., 2015) known today as Extreme Gradient Boosting (XGBoost) algorithm. It has been reported (Alessandretti, ElBahrawy, Aiello, & Baronchelli, 2018, p. 4) this method has been used in a number of winning Kaggle solutions (17/29 in 2015). XGBoost has been applied to the Bitcoin market in (Chen, Li, & Sun, 2020, p. 12) and its accuracy has been reported 0.483. Another experiment on XGB-based methods has yielded $1.1 * 10^3$ BTC (for their method 1) and $\sim 95$ BTC (for their Method 2) starting from 1 BTC (Alessandretti, ElBahrawy, Aiello, & Baronchelli, 2018, p. 7)

### III. METHODS AND MATERIALS

In this section we first look at the data used in this project, then we get acquainted with three different methods which have been used to make the models for the prediction task. We have tested other machine learning methods in our settings including: Ridge Classification (Rifkin, Yeo, & Poggio, 2003), Logistic Regression (LaValley, 2008), Stochastic Gradient Descent (Kabir, Siddique, Alam Kotwal, & Nurul Huda, 2015), Multi Layer Perceptron (Pal & Mitra, 1992), Support Vector Machines (Hearst, Dumais, Osuna, Platt, & Scholkopf, 1998), Gaussian Process Classification (Csató, Fokoué, Opper, Schottky, & Winther, 1999), Gaussian Naïve Bayes (Webb, Keogh, & Miikkulainen, 2010) and Decision Trees (Rokach & Maimon, 2005). Our best results were with the following methods, other methods yielded negative profits or lesser profits relative to the following methods (less than half of the profit factors on average).

1. *Used Data*

Binance is a cryptocurrency exchange that provides a platform for trading various cryptocurrencies. As of April 2021, Binance was the largest cryptocurrency exchange in the world in terms of trading volume (Top Cryptocurrency Spot Exchanges, 2021).

Binance provides a free to use API for data gathering. This API is conveniently available to use in python (Patro & Sahu, 2015). We use this API to gather Time stamp (in second precision), Open, High, Low, Close and Volume for a 4 hours period dataframe. This procedure runs for all three different assets that we study: ETH-USDT, LTC-BTC and

ZEC-BTC. Data gets gathered from mid-2017 until April 2021. This makes our raw input data.

2. *First Classifier: k-Nearest Neighbours Vote*

Neighbours-based models are type of *instance-based learning* or *non-generalizing learning*. They don't attempt to construct a general internal model, but simply store instances of the training data (hence called *lazy learners)*. Classification is computed from a sort of majority vote of the nearest neighbours of each point: the point we are trying to classify is assigned to the data class which has the most representatives within the nearest neighbours of the point. Using distance metrics can sometimes improve the accuracy of the model. (Pedregosa, et al., 2011) These models are also beneficial for regression problems.

Suppose we have pairs $(X_1, Y_1), (X_2, Y_2),\ldots, (X_n, Y_n)$ taking values in $R^d \times \{1,2\}$, where $Y$ is the class label of $X$, so that $X|Y = r \sim P_r$ for $r = 1,2$ (and probability distributions $P_r$). Given some norm $||.||$ on $R^d$ and a point $x \in R^d$, let $(X_{(1)}, Y_{(1)}), (X_{(2)}, Y_{(2)}), \ldots, (X_{(n)}, Y_{(n)})$ be a reordering of the training data such that $||X_{(1)} - x|| \leq \ldots ||X_{(n)} - x||$. Now, by voting on $X_{(i)}$ starting from $i = 1$ and going increasingly for $i$, we can do the classification task. (Cannings, Berrett, & Samworth, 2020)

We use Scikit-learn implementation (Pedregosa, et al., 2011) of k-NN classifier in this project.

3. *Second Classifier: Random Forest*

Random forests or random decision forests are classified as ensemble learning methods. They can be applied to classification, regression and other tasks that operate by constructing an assembly of decision trees at training time and returning the class that is the mode of the classes (for classification) or mean/average prediction (for regression) of the individual trees (Ho, Random decision forests, 1995). "Random decision forests correct for decision trees' habit of over fitting to their training set" (Hastie, Tibshirani, & Friedman, 2009, pp. 587-588). Random forests generally perform better than individually assisted decision trees, but their accuracy could be lower than gradient boosted trees. However, data characteristics can affect their performance (Piryonesi & El-Diraby, Data Analytics in Asset Management: Cost-Effective Prediction of the Pavement Condition Index, 2019) (Piryonesi & El-Diraby, Role of Data Analytics in Infrastructure Asset Management: Overcoming Data Size and Quality Problems, 2020).

The training algorithm for random forests uses the general technique of bootstrap aggregating, or bagging, to tree learners. Given a training set $X = x_1,\ldots,x_n$ with labels $Y = y_1,\ldots,y_n$ bagging repeatedly ($B$ times) selects a random sample with replacement of the training set and fits trees to these samples:

for b=1, … , B:
I. Sample, with replacement, n training examples from $X,Y$; call these $X_b,Y_b$.
II. Train a classification or regression tree $f_b$ on $X_b,Y_b$.

After training, prediction for unseen sample $x'$ can be made by averaging the predictions from all the individual regression trees on $x'$:

$$\hat{f} = \frac{1}{B}\sum_{b=1}^{B} f_b(x')$$

or by taking the majority vote in the case of classification trees. (Lu, 2014)

This bootstrapping procedure results in better model performance because it decreases the variance of the model, without increasing the bias. This means that while the predictions of a single tree are highly susceptible to noise in its training set, the average of many trees is not, as long as the trees are not correlated. Simply training many trees on a single training set will produce strongly correlated trees (or even the same tree many times, if the training algorithm is deterministic); bootstrap sampling is a way to de-correlate the trees by providing them different training sets. (Lu, 2014)

The number of samples/trees, $B$, is a free parameter. Typically, a few hundred to several thousand trees are used, depending on the size and nature of the training set. An optimal number of trees, $B$, can be found using cross-validation, or by observing the out-of-bag error: the mean prediction error on each training sample $x_i$, using only the trees that did not have $x_i$ in their bootstrap sample (James, Witten, Hastie, & Tibshirani, 2013, pp. 316-321).

We use Scikit-learn implementation (Pedregosa, et al., 2011) of random forest classifier in this project.

## 4. Third Classifier: eXtreme Gradient Boosting

Gradient boosting is a machine learning technique for regression and classification problems, which like random forest, produces a prediction model in the form of an ensemble of weak prediction models, typically decision trees. When a decision tree is the weak learner, the resulting algorithm is called gradient boosted trees, which usually outperforms random forest (Piryonesi & El-Diraby, Using Machine Learning to Examine Impact of Type of Performance Indicator on Flexible Pavement Deterioration Modeling, 2021) (Friedman, Tibshirani, & Hastie, 2009). It builds the model in a stage-wise fashion like other boosting methods do, and it generalizes them by allowing optimization of an arbitrary differentiable loss function.

Like other boosting methods, gradient boosting combines weak "learners" into a single strong learner in an iterative fashion. It is easiest to explain in the least-squares regression setting, where the goal is to "teach" a model $F$ to predict values of the form $\hat{y} = F(x)$ by minimizing the mean squared error $\frac{1}{n}\sum_i(\hat{y}_i - y_i)^2$ where $i$ indexes over some training set of size $n$ of actual values of the output variable $y$. So we have following definition:

$\hat{y}_i$ = The Predicted Value $F(x)$
$y_i$ = The Observed Value
$n$ = The Number of Samples in $y$

Now, let us consider a gradient boosting algorithm with $M$ stages. At each stage $m$ $(1 \leq m \leq M)$ of gradient boosting, suppose some imperfect model $F_m$ (for low $m$ this model may simply return $\hat{y}_i = \bar{y}$, where the right-hand side is the mean of $y$). In order to improve $F_m$, our algorithm should add some new estimator, $h_m(x)$. Thus,

$$F_{m+1}(x) = F_m(x) + h_m(x) = y$$

or, equivalently,

$$h_m(x) = y - F_m(x).$$

Therefore, gradient boosting will fit $h$ to the residual $y - F_m(x)$. As in other boosting variants, each $F_{m+1}$ attempts to correct the errors of its predecessor $F_m$. A generalization of this idea to loss functions other than squared error, and to classification and ranking problems, follows from the observation that residuals $h_m(x)$ for a given model are the negative gradients of the mean squared error (MSE) loss function (with respect to $F(x)$):

$$L_{MSE} = \frac{1}{2}(y - F(x))^2$$
$$h_m(x) = -\frac{\partial L_{MSE}}{\partial F} = y - F(x).$$

So, gradient boosting could be specialized to a gradient descent algorithm, and generalizing it entails "plugging in" a different loss and its gradient (Li, 2021).

Now, with having an overview of boosted trees, one may ask what are XGBoost trees? XGBoost is a tool motivated by the formal principle introduced. More importantly, "it is developed with both deep consideration in terms of systems optimization and principles in machine learning". "The goal of this library is to push the extreme of the computation limits of machines to provide a scalable, portable and accurate library" (Chen & Guestrin, Xgboost: A scalable tree boosting system, 2016). We use this library for our implementations of the solution.

## IV. PROPOSED METHODS

In this section we look at our proposed methods for market direction problem in cryptocurrency markets. First, we fill in details about our raw data gathering procedure. Then at the second subsection, we elaborate on our pre-processing steps for the obtained raw financial data. We also explain the dataset creation part of the scheme at this subsection. The third step, sums up the definition of our three different models. We also make some concise points about hyperparameters of each model. Last subsection looks at evaluation of results and concludes the strategy design part of the system. Figure 1 (look at page 8) shows a comprehensive view of the whole system, green lines indicate train phase and red lines indicate exertion phase.

## 1. Raw Data Gathering

At the time of doing this research, Binance has made available, access to its historical records (for Open, High, Low, Close and Volume) through its API for time frames bigger than one minute. We gather 4 hour period data to a Pandas dataframe since its first available timestamp (which is usually mid 2017). The data includes OHLCV for ETH-USDT, LTC-BTC, ZEC-BTC. As cryptocurrency markets are very dynamic environments (price and return levels can be changed dramatically just in some months) we opt to have as much as recent data as we can for the current state of the market, so we decided to use 95% of the data for training and the remaining 5% to evaluate the models (which makes the test data time span almost 66 days). In practical application of this system we can also retrain the models in a period shorter than this 66 days' time span.

2. *Pre Processing and Dataset Creation*

After gathering the data, we make two copies of it. For one copy, we augment it with some famous Technical Indicators in finance. Essentially these indicators are some mathematical functions which take some arguments from real time or past data and they create some insights about "technical" moves of the market. Name and formula for these technical indicators have been reported in the Appendix A. The other copy doesn't include these technical indicators data.

After augmentation part we have two dataframes including all relevant data for each record. Now, two things must be done to make our datasets a suitable one for our models: 1- We need to encapsulate all the features used for identifying a data point for each one of them, 2- We need to label each datapoint.

For this project we use the last financial records plus 59 records proceeding it as its features. These records are in a 4 hour period and all of the produced features are numerical. We normalize them by using the Feature Scaling method (Patro & Sahu, 2015). Each value gets divided to its maximum value minus its minimum.

To encapsulate the feature data for each datapoint we take all the 60 rows (and 19 parameters for augmented version and 5 parameters for unaugmented version at each row) from our dataframes and put all those variables inside another array named X. So, each $X_i$ will be a datapoint with 1140 parameters for augmented version and with 300 parameters for unaugmented version.

To label each datapoint we define a threshold to determine if retrospectively we would have entered the market at that timestamp, after 4 hours we would make profit or loss? This threshold gets defined using the fees per each cryptocurrency exchanger. At the time of doing this research the least possible fee to exchange in the market in Binance was 0.15 percent (0.075 percent to exchange from A symbol to B symbol and 0.075 to exchange from B to original A symbol). So, we define our threshold in this project as about 0.15 percent of the price movement in the 4 hour period. To sum up, if an asset's value changes more than this threshold in a positive direction, we label that as "1" and otherwise we label it as "0", This way per any label=1 if we had entered the market at that point we would make profit.

3. *Model definition- Model Training*

At this subsection we look at the libraries and hyperparameters involved in each model. We also note each model's training time. A more elaborate discussion about the hyper parameters is held in the Discussion section.

kNN model and random forest have been implemented using open source machine learning library Scikit-learn. Scikit-learn features various classification, regression and clustering algorithms including support vector machines, random forests, gradient boosting, k-means and DBSCAN, and is designed to interoperate with the Python numerical and scientific libraries NumPy and SciPy (Pedregosa, et al., 2011). An important design note about scikit-learn is its unified interface for its models. If user's data suffices this interface requirements, it's easy to use and change models to use for the same data.

XGB has been implemented using XGBoost. XGBoost is an optimized distributed gradient boosting library designed to be highly efficient, flexible and portable. It implements machine learning algorithms under the Gradient Boosting framework. XGBoost provides a parallel tree boosting (also known as GBDT, GBM) that solve many data science problems in a fast and accurate way (Chen & Guestrin, Xgboost: A scalable tree boosting system, 2016).

Hyperparameters involved in kNN classifier are as follows:

*Number of Neighbours:* Depended on The Dataset (5 for ETHUSDT, 20 for LTCBTC, 100 for ZECBTC)

*Weight Function Used in Prediction:* Distance

*Algorithm Used to Compute The Nearest Neighbours:* Auto: will attempt to decide the most appropriate algorithm between BallTree, KDTree and Brute Force based on the values passed to fit method.

*Leaf Size:* 30

*The Distance Metric to Use for The Tree:* Minkowski

*Power Parameter for The Minkowski Metric:* 2

Hyperparameters involved in Random Forest classifier are as follows:

*The number of trees in the forest:* Depended on The Dataset (700 for ETHUSDT and ZECBTC, 1000 for LTCBTC)

*The Function to Measure The Quality of A Split:* gini

*The Maximum Depth of The Tree:* Nodes are expanded until all leaves are pure or until all leaves contain less than the minimum number of samples required to split an internal node samples.

*The minimum number of samples required to split an internal node:* 2

Hyperparameters involved in XGB classifier are as follows:

*Booster:* gbtree

*Eta (alias: Learning Rate):* 0.3

*Minimum Loss Reduction Required to Make A Further Partition on A Leaf Node of The Tree (The larger gamma is, the more conservative the algorithm will be.):* 0

*Maximum Depth of A Tree:* 6

*Lambda (L2 regularization term on weights. Increasing this value will make model more conservative.):* 1

*Alpha (L1 regularization term on weights. Increasing this value will make model more conservative.):* 0

Training and evaluation of the models in this project has been done using Colab virtual machines by google. Training time takes the most for Random Forest with an average of 167.97 seconds. Second place goes to XGB with an average of 46.85 seconds and finally kNN takes only 1.06 seconds on average to be trained for these datasets.

*4. Evaluation and Strategy Design*

To evaluate each model in this project we use two different methods: Accuracy of The Model and The Profit Obtained by The Model. By accuracy in this context, we mean how many times the predicted label for the market direction matches with the real direction of the market. To discuss how we calculate the obtained profit, we need to understand how we use the models to create the strategy.

Strategy Design procedure is pretty straightforward. We took 60 rows of records from now and past data and then we decide if the price will go up enough to cover the exchange's fee? If we predict it will, we enter the market and after 4 hours we retry this operation to decide for the next 4 hours. If the next 4 hours still shows an adequate positive movement, we keep the buy position and if it does not, we sell what we have bought. Now, our profit is the difference between the values of bought and sold assets. It will accumulate this positive or negative profit through the test span. Notice that our position size stays the same at each buying or selling trade. At the final step of the strategy we sell whatever we have. Another evaluative indicator for strategy assessments in financial markets is Profit Factor, which is defined as: "The gross profit divided by the gross loss (including commissions) for the entire trading period". We calculate this metric for each model and each asset. In the next section we look at the results of our experiments with our models.

V. EXPERIMENTAL RESULTS

Here we look at three different cryptocurrencies that we study, separately. This section has three subsections relative to each cryptocurrency pair. At each subsection, first, we have a graph of the pair's price movements through the time span that we scrutinize it. Then, we have a graph that shows the normalized returns of that pair through the time. The mentioned graph shows what we are trying to predict. The third graph is the pair's price movements through its test phase. After that, we have performance graphs for both augmented and original input data. Performance is assessed with a cumulative reward graph which shows how much money we have earned with a fixed position at each time we entered or exited the market. Finally, we have some information regarding the test of models and a distribution graph of each model's (positive or negative) profits. For the sake of concision, for the unaugmented experiments we just report the performance graphs.

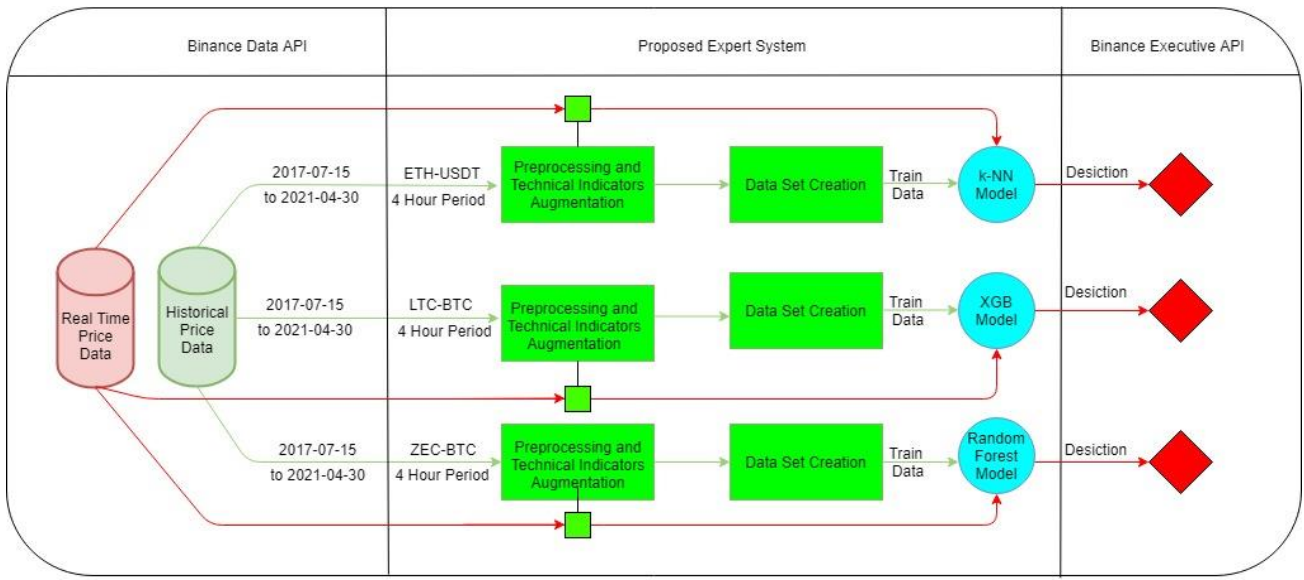

Figure 1. Overall Structure of The Proposed Expert System. Green lines indicate train phase and red lines indicate exertion phase

*ETH-USDT:*

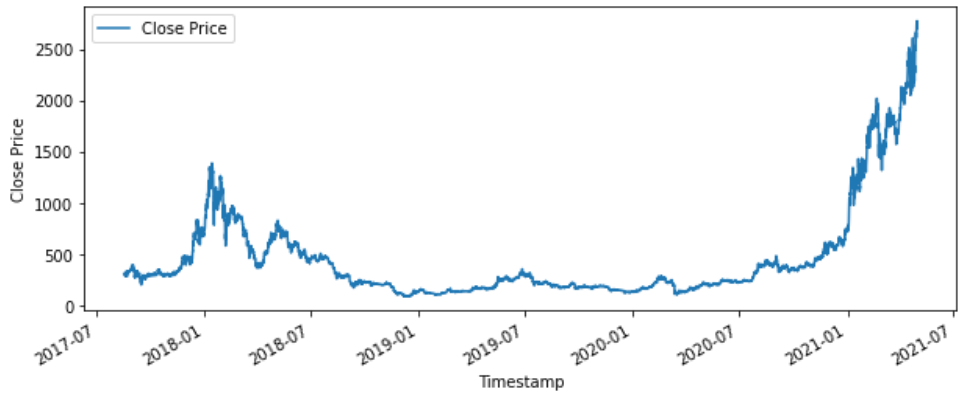

Figure 2. Close Price for ETH-USDT from 2017-07 to 2021-05

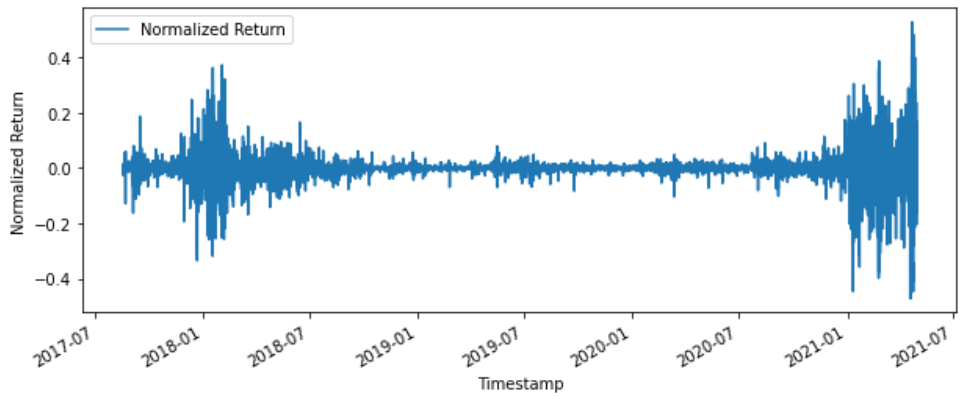

Figure 3. Normalized Return for ETH-USDT from 2017-07 to 2021-07

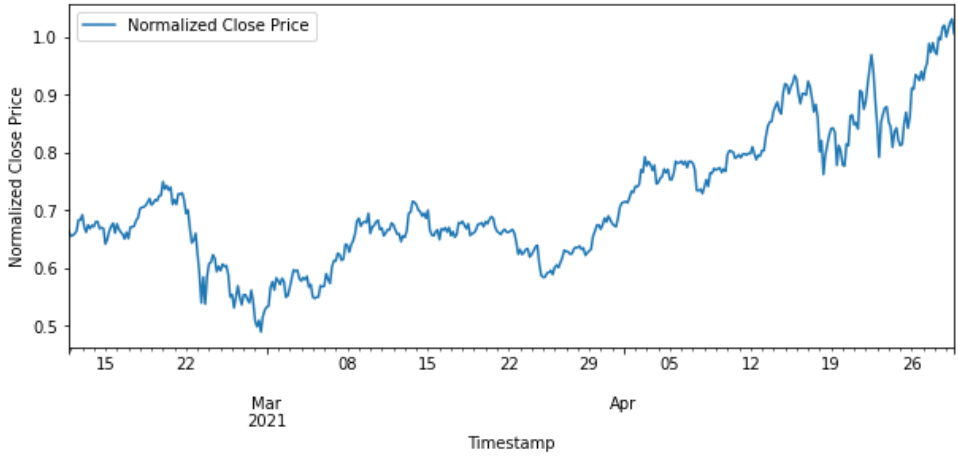

Figure 3. Normalized Close Price for ETH-USDT in Test Data

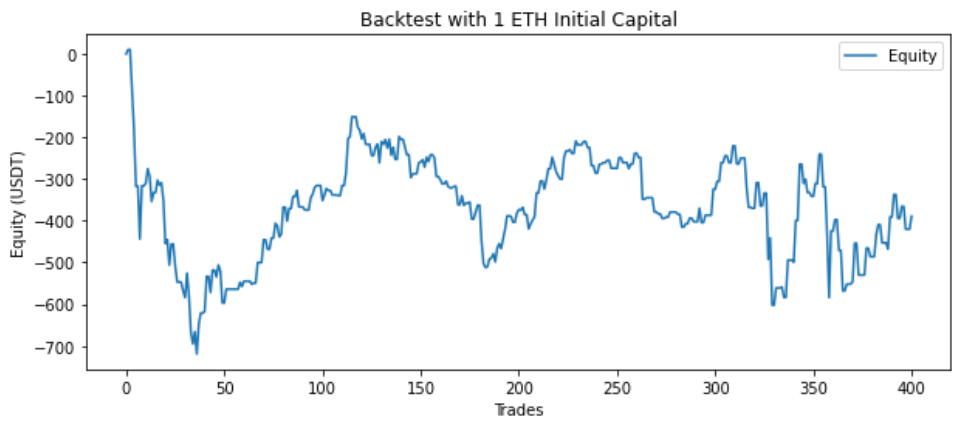

Figure 4. Performance of The k-NN Model for ETH-USDT in Unaugmented Test Data

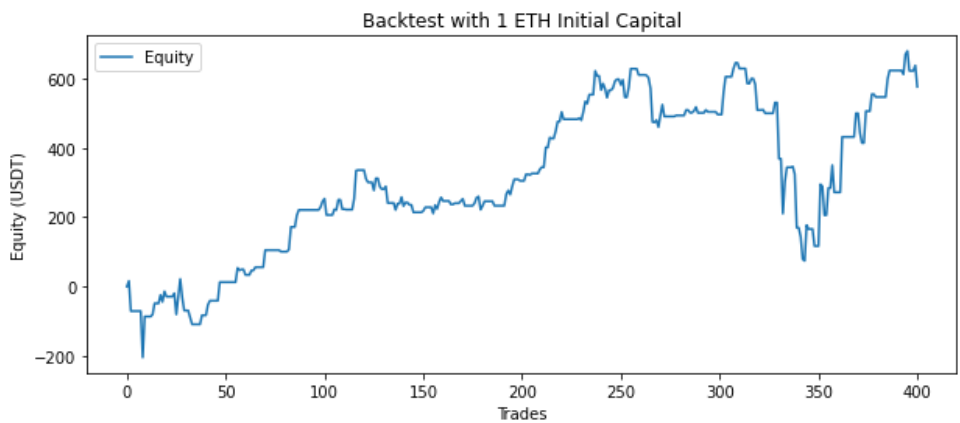

Figure 5. Performance of The k-NN Model for ETH-USDT in Augmented Test Data

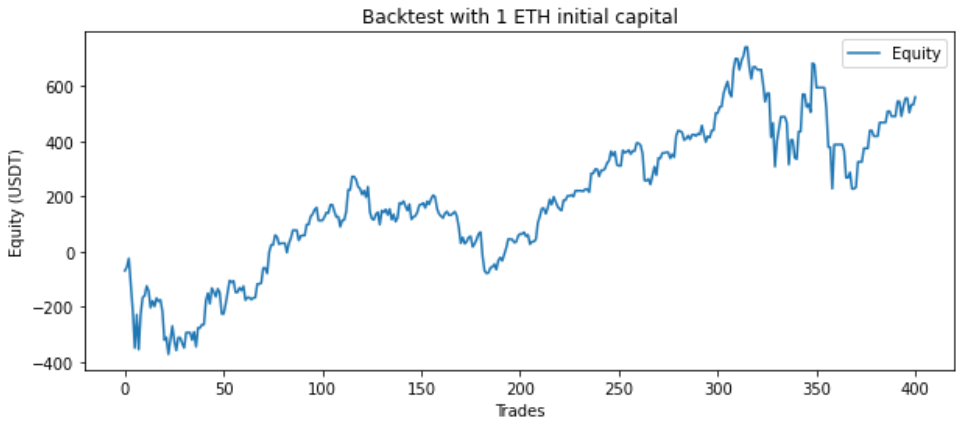

Figure 6. Performance of The RF Model for ETH-USDT in Unaugmented Test Data

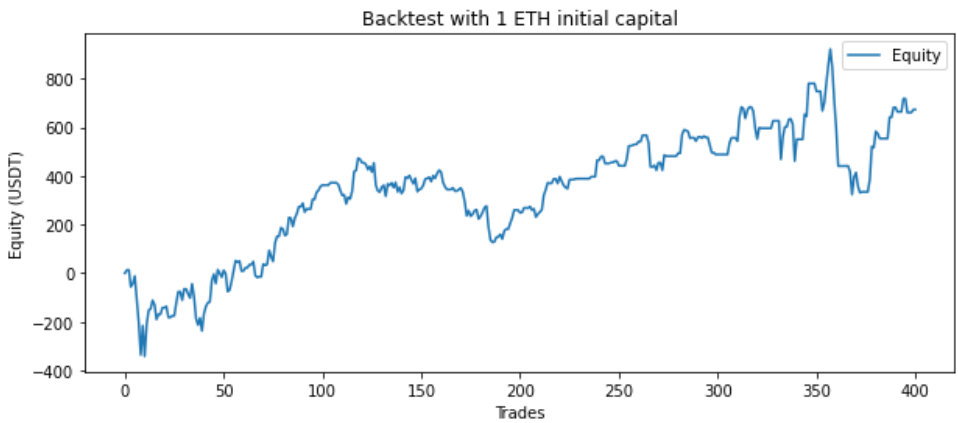

Figure 7. Performance of The RF Model for ETH-USDT in Augmented Test Data

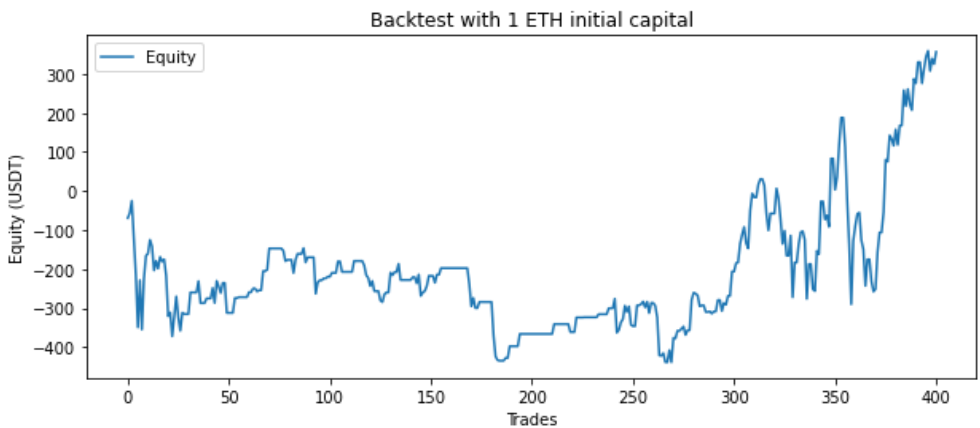

Figure 8. Performance of The XGB Model for ETH-USDT in Unaugmented Test Data

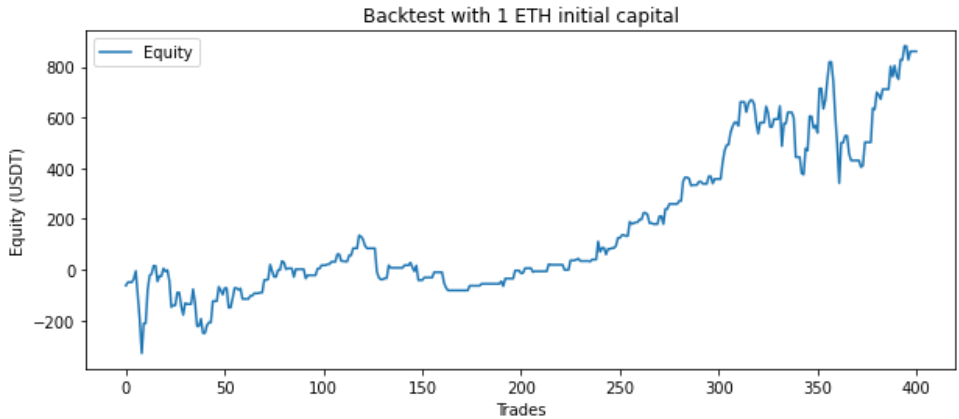

Figure 9. Performance of The XGB Model for ETH-USDT in Augmented Test Data

| Testing Accuracy | 0.519900 |
|---|---|
| Net Profit | 575.810 |
| Number of Winning Trades | 105 |
| Number of Losing Trades | 82 |
| Total Days in Test | 66 |
| Percent of Profitable Trades | 56.15% |
| Avg Win Trade | 29.680 |
| Avg Los Trade | -30.983 |
| Largest Win Trade | 177.820 |
| Largest Los Trade | -161.700 |
| Profit Factor | 1.23 |

Table 1. Information Regarding k-NN Test on ETH-USDT

| Testing Accuracy | 0.562189 |
|---|---|
| Net Profit | 672.80 |
| Number of Winning Trades | 166 |
| Number of Losing Trades | 125 |
| Total Days in Test | 66 |
| Percent of Profitable Trades | 57.04% |
| Avg Win Trade | 29.782 |
| Avg Los Trade | -34.168 |
| Largest Win Trade | 135.050 |
| Largest Los Trade | -158.100 |
| Profit Factor | 1.16 |

Table 2. Information Regarding RF Test on ETH-USDT

| | |
|---|---|
| Testing Accuracy | 0.547264 |
| Net Profit | 860.940 |
| Number of Winning Trades | 120 |
| Number of Losing Trades | 90 |
| Total Days in Test | 66 |
| Percent of Profitable Trades | 57.14% |
| Avg Win Trade | 36.302 |
| Avg Los Trade | -38.836 |
| Largest Win Trade | 174.820 |
| Largest Los Trade | -158.100 |
| Profit Factor | 1.25 |

Table 3. Information Regarding XGB Test on ETH-USDT

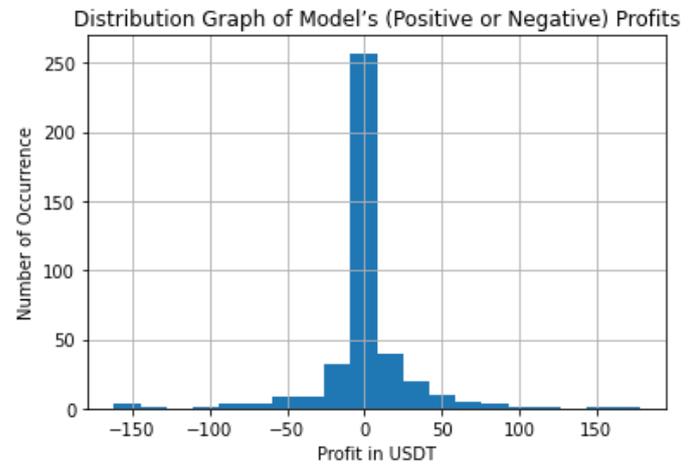

Figure 10. Distribution of Profits for k-NN in ETH-USDT

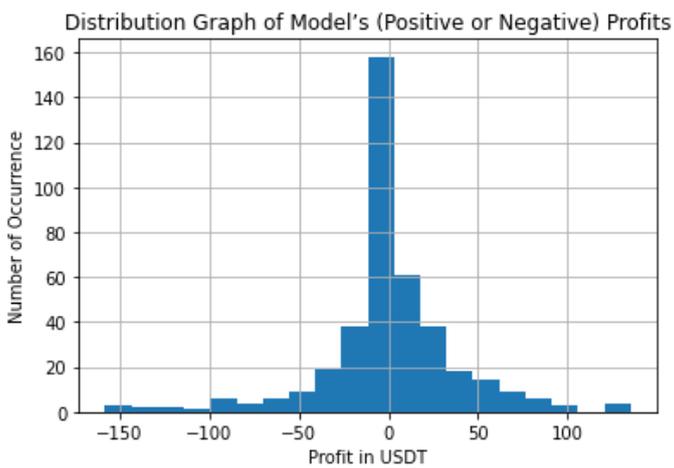

Figure 11. Distribution of Profits for RF in ETH-USDT

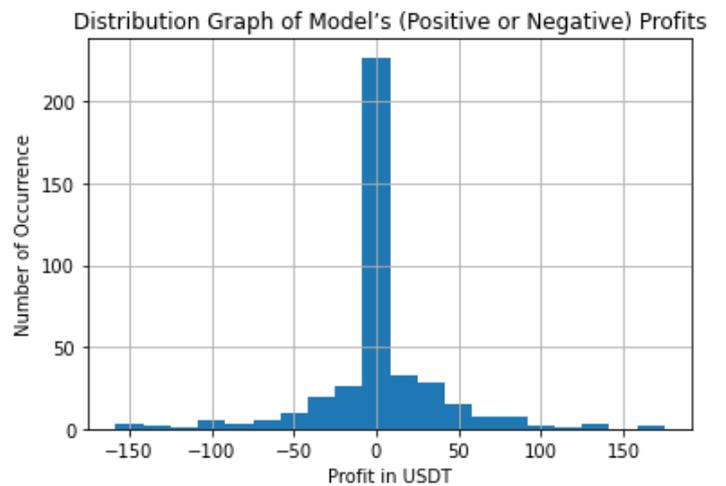

Figure 12. Distribution of Profits for XGB in ETH-USDT

*LTC-BTC:*

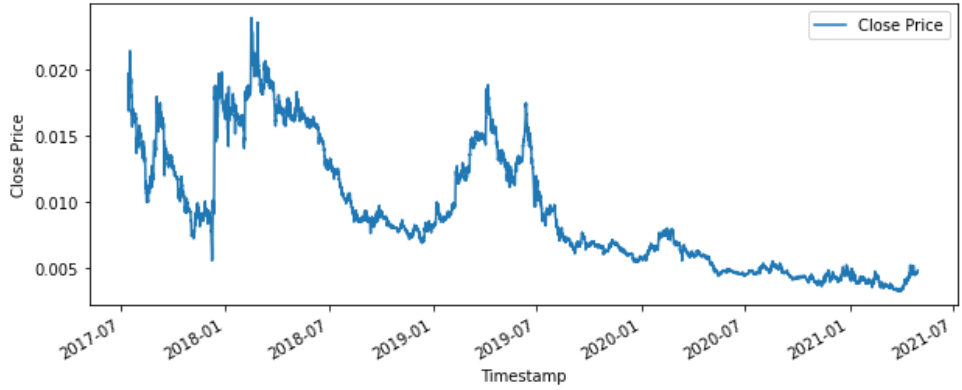

Figure 13. Close Price for LTC-BTC from 2017-07 to 2021-07

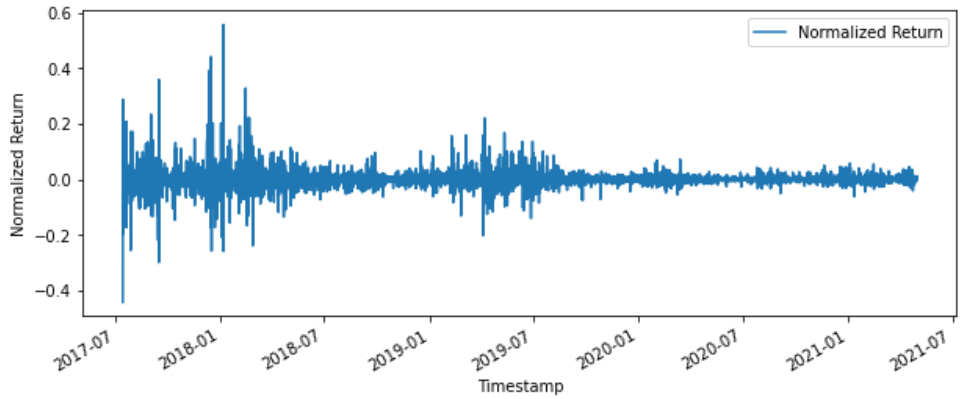

Figure 14. Normalized Return for LTC-BTC from 2017-07 to 2021-07

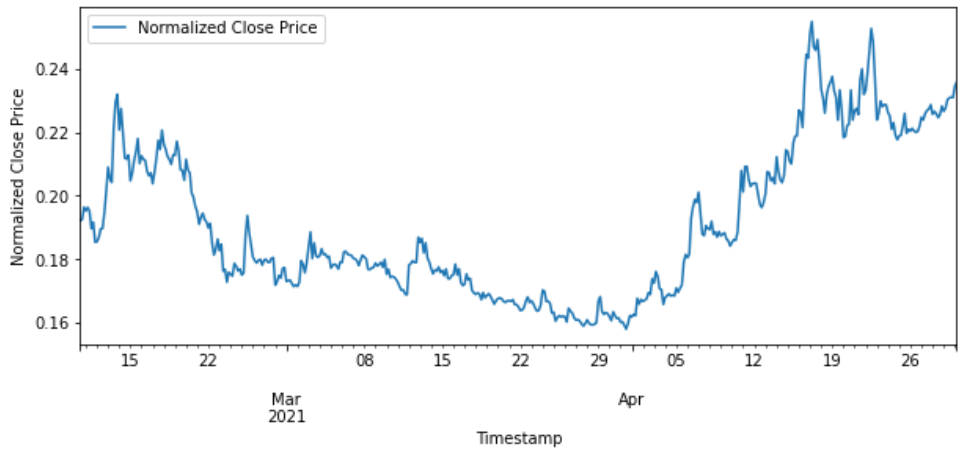

Figure 15. Normalized Close Price for LTC-BTC in Test Data

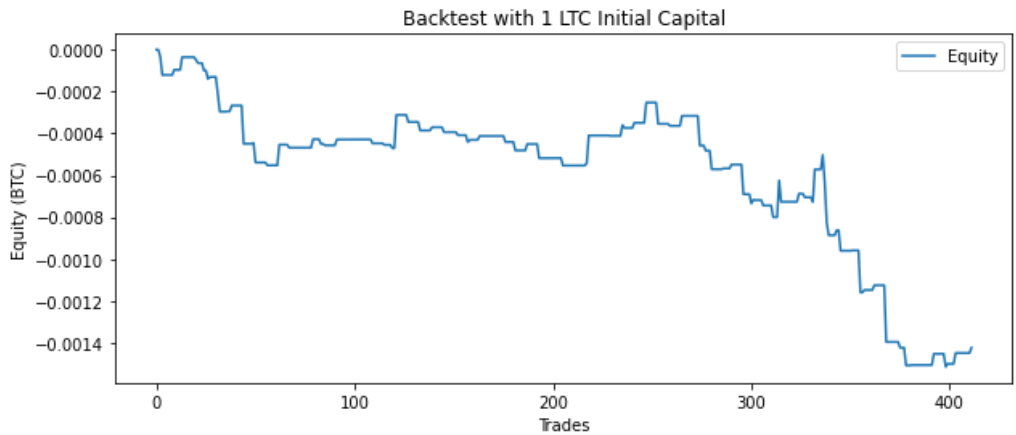

Figure 16. Performance of The k-NN Model for LTC-BTC in Unaugmented Test Data

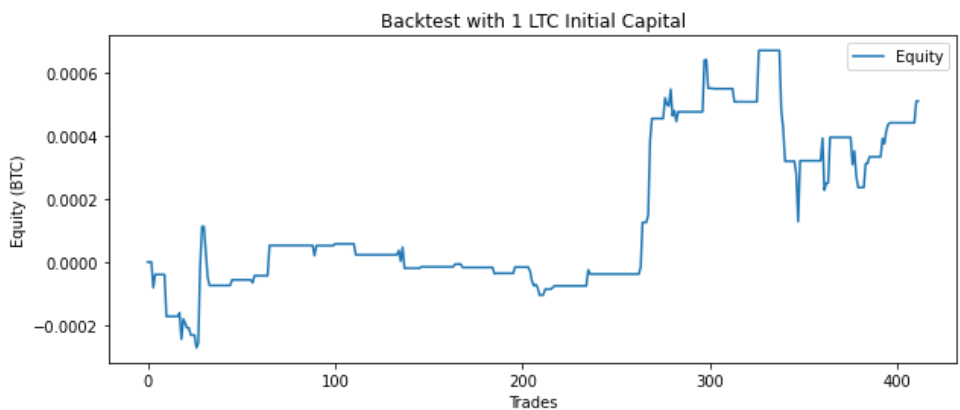

Figure 17. Performance of The k-NN Model for LTC-BTC in Augmented Test Data

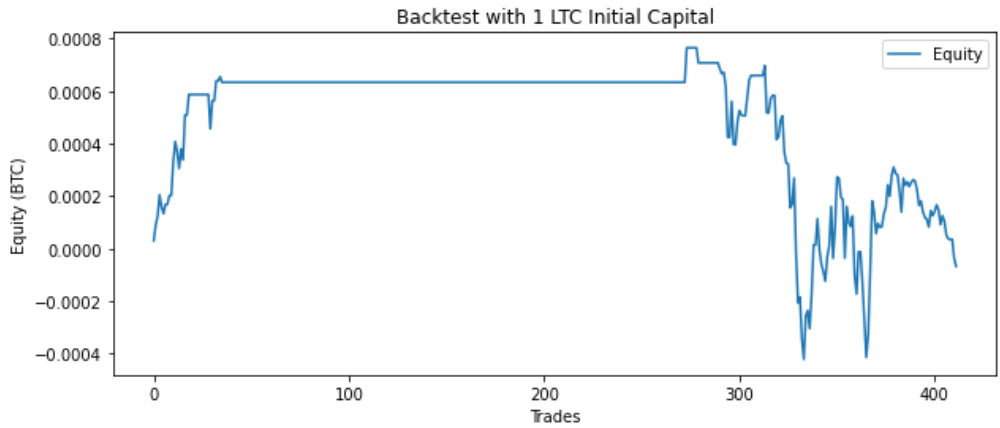

Figure 18. Performance of The RF Model for LTC-BTC in Unaugmented Test Data

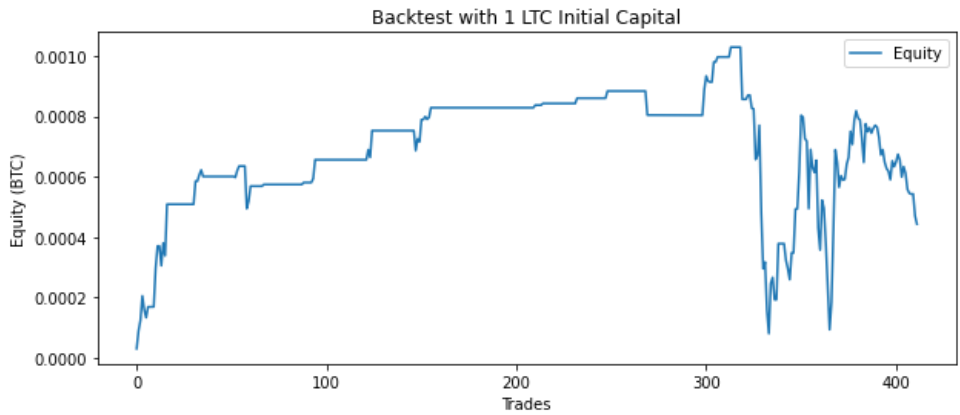

Figure 19. Performance of The RF Model for LTC-BTC in Augmented Test Data

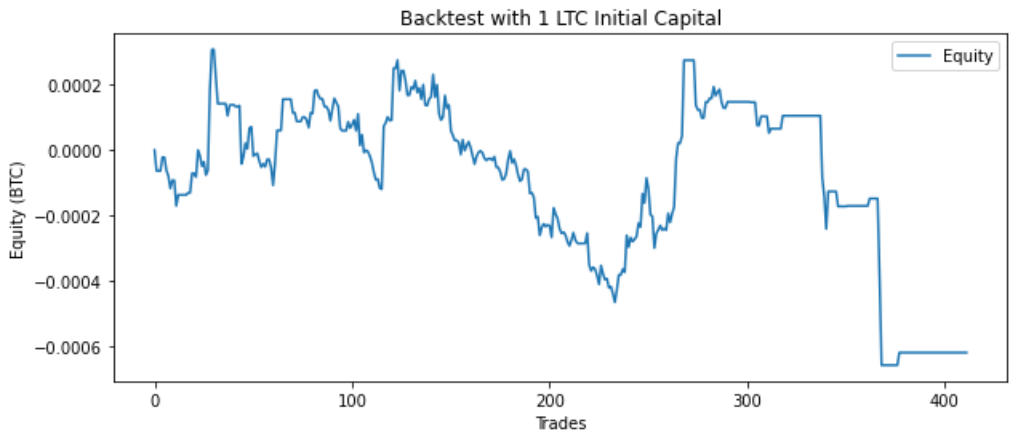

Figure 20. Performance of The XGB Model for LTC-BTC in Unaugmented Test Data

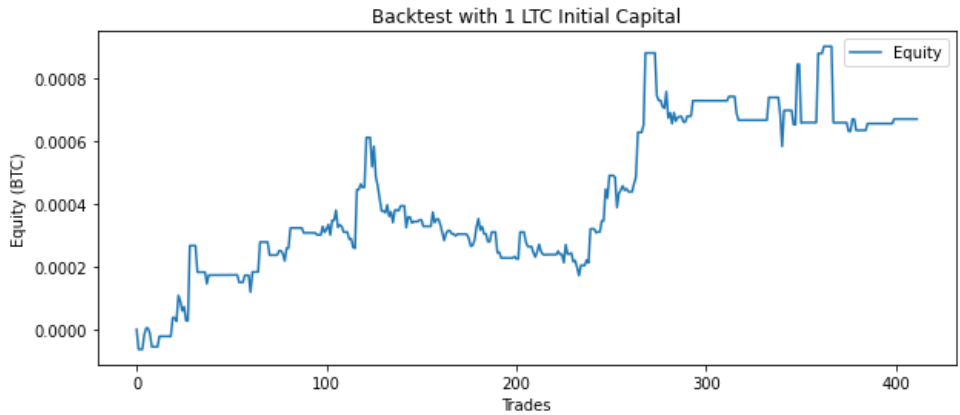

Figure 21. Performance of The XGB Model for LTC-BTC in Augmented Test Data

| | |
|---|---|
| Testing Accuracy | 0.585956 |
| Net Profit | 0.0005090 |
| Number of Winning Trades | 46 |
| Number of Losing Trades | 40 |
| Total Days in Test | 66 |
| Percent of Profitable Trades | 53.49% |
| Avg Win Trade | 0.00006 |
| Avg Los Trade | -0.00005 |
| Largest Win Trade | 0.00024 |
| Largest Los Trade | -0.00019 |
| Profit Factor | 1.24 |

Table 4. Information Regarding k-NN Test on LTC-BTC

| | |
|---|---|
| Testing Accuracy | 0.467312 |
| Net Profit | 0.0004430 |
| Number of Winning Trades | 71 |
| Number of Losing Trades | 65 |
| Total Days in Test | 66 |
| Percent of Profitable Trades | 52.21% |
| Avg Win Trade | 0.00006 |
| Avg Los Trade | -0.00006 |
| Largest Win Trade | 0.00027 |
| Largest Los Trade | -0.00029 |
| Profit Factor | 1.12 |

Table 5. Information Regarding RF Test on LTC-BTC

| | |
|---|---|
| Testing Accuracy | 0.520581 |
| Net Profit | 0.0006720 |
| Number of Winning Trades | 88 |
| Number of Losing Trades | 91 |
| Total Days in Test | 66 |
| Percent of Profitable Trades | 49.16% |
| Avg Win Trade | 0.00004 |
| Avg Los Trade | -0.00003 |
| Largest Win Trade | 0.00024 |
| Largest Los Trade | -0.00024 |
| Profit Factor | 1.22 |

Table 6. Information Regarding XGB Test on LTC-BTC

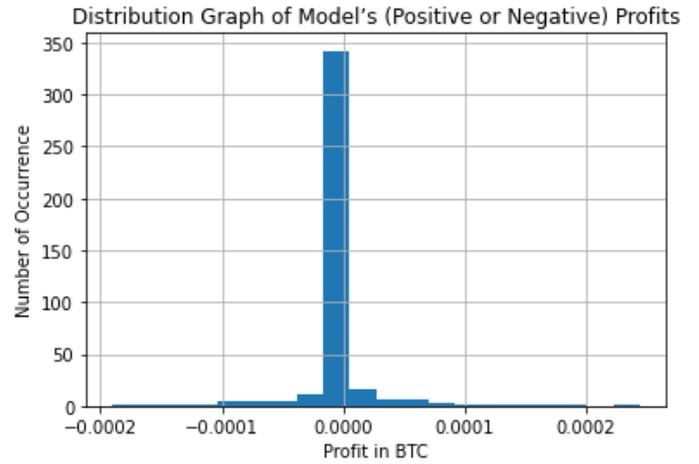

Figure 22. Distribution of Profits for k-NN in LTC-BTC

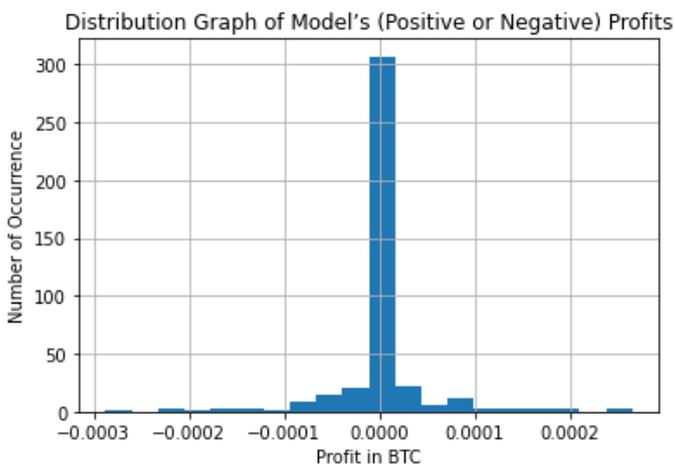

Figure 23. Distribution of Profits for RF in LTC-BTC

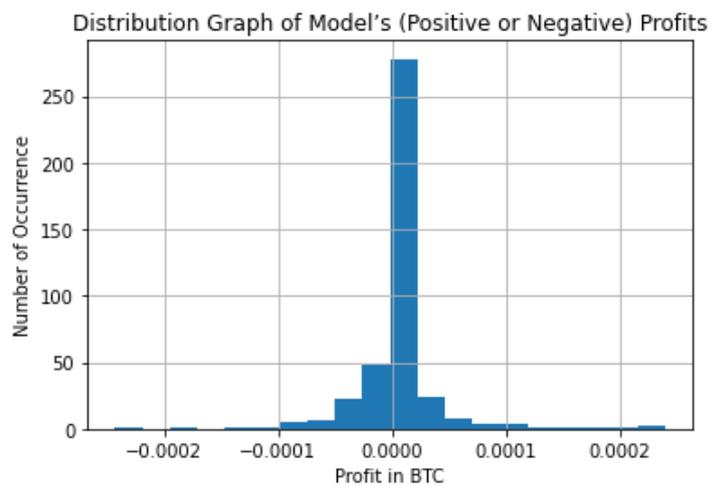

Figure 24. Distribution of Profits for XGB in LTC-BTC

*ZEC-BTC:*

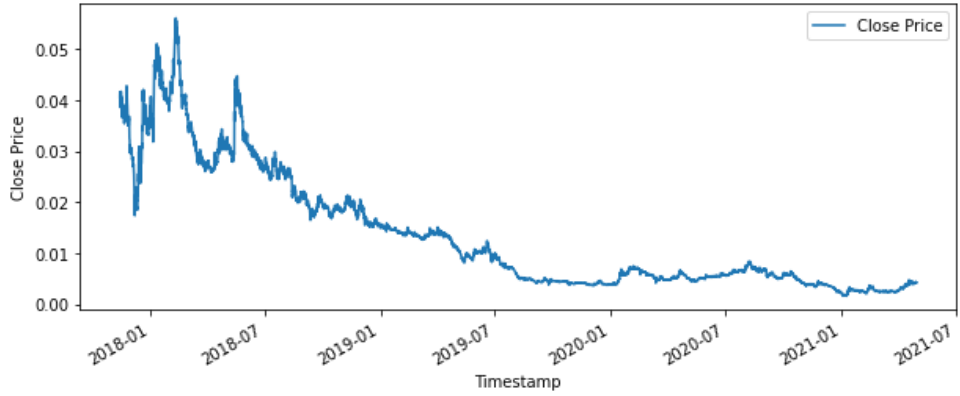

Figure 25. Close Price for ZEC-BTC from 2017-07 to 2021-07

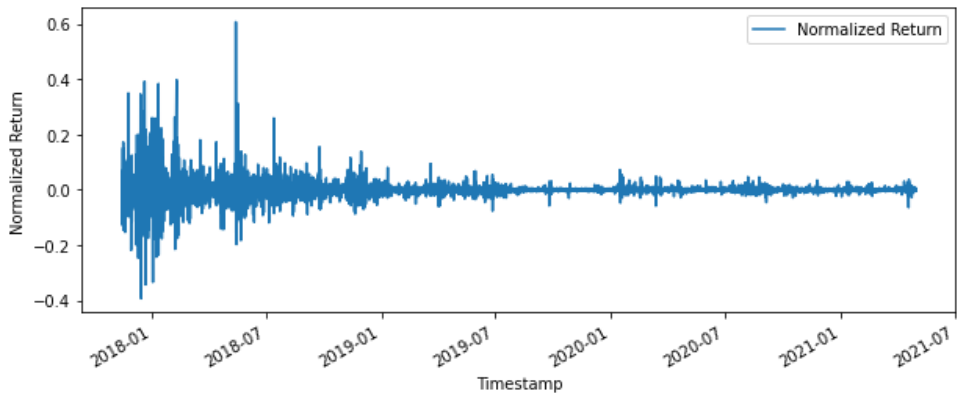

Figure 26. Normalized Close Price for ZEC-BTC in Test Data

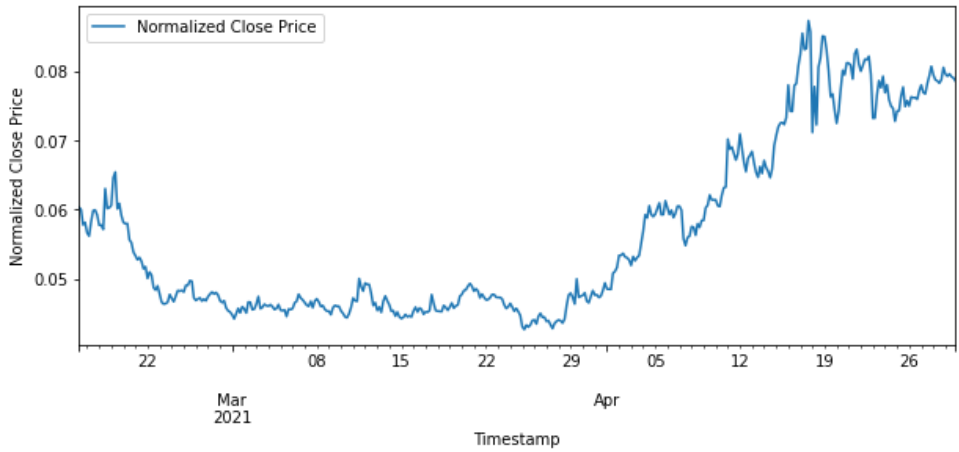

Figure 27. Normalized Close Price for ZEC-BTC in Test Data

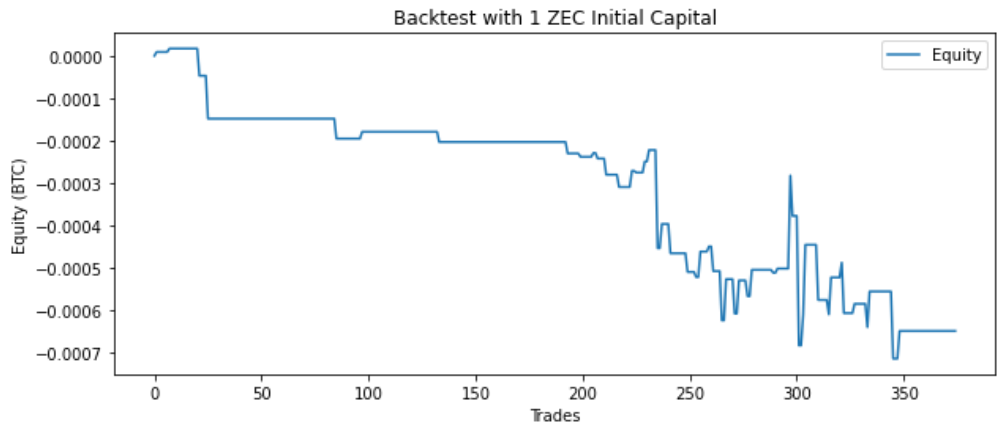

Figure 28. Performance of The k-NN Model for ZEC-BTC in Unaugmented Test Data

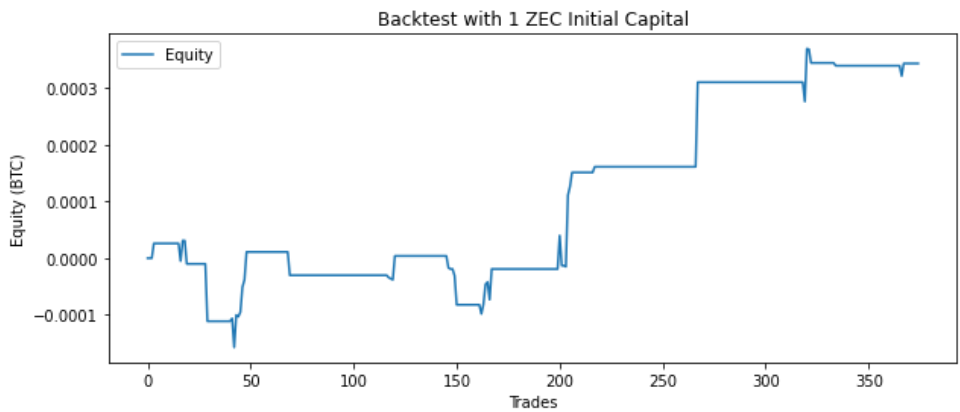

Figure 29. Performance of The k-NN Model for ZEC-BTC in Augmented Test Data

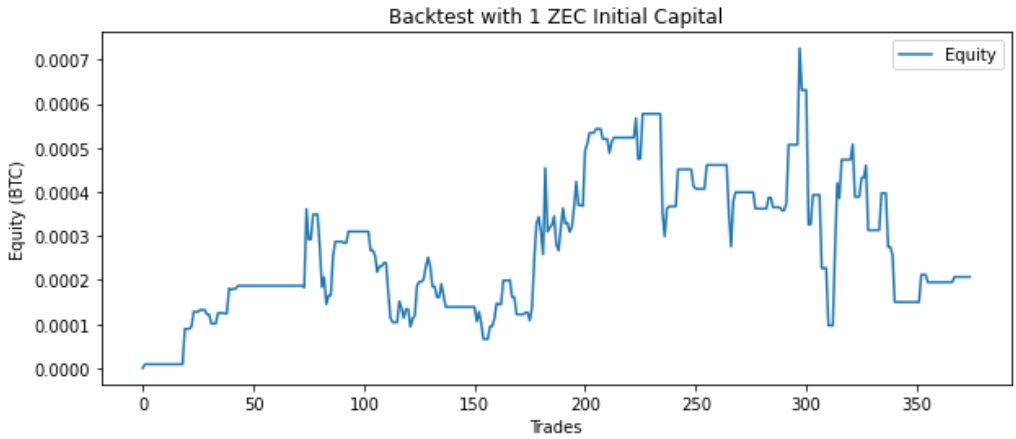

Figure 30. Performance of The RF Model for ZEC-BTC in Unaugmented Test Data

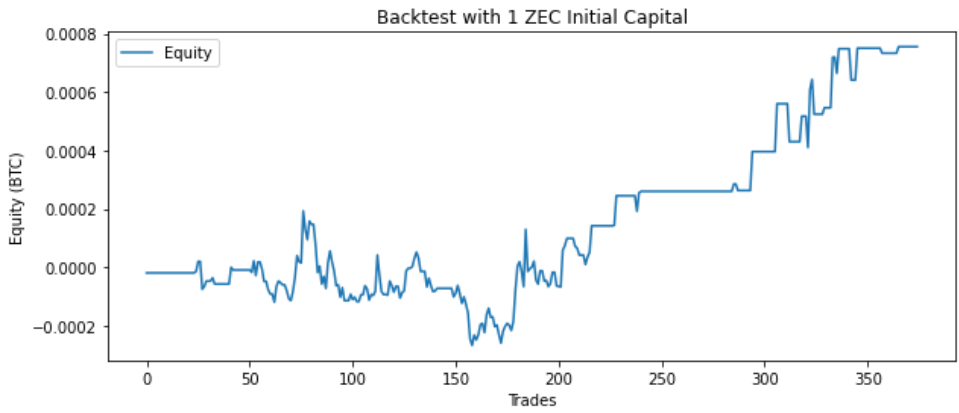

Figure 31. Performance of The RF Model for ZEC-BTC in Augmented Test Data

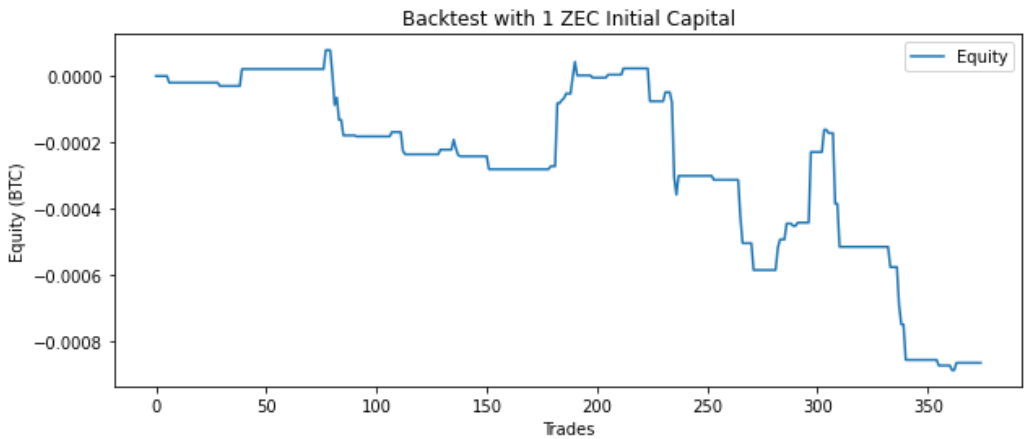

Figure 32. Performance of The XGB Model for ZEC-BTC in Unaugmented Test Data

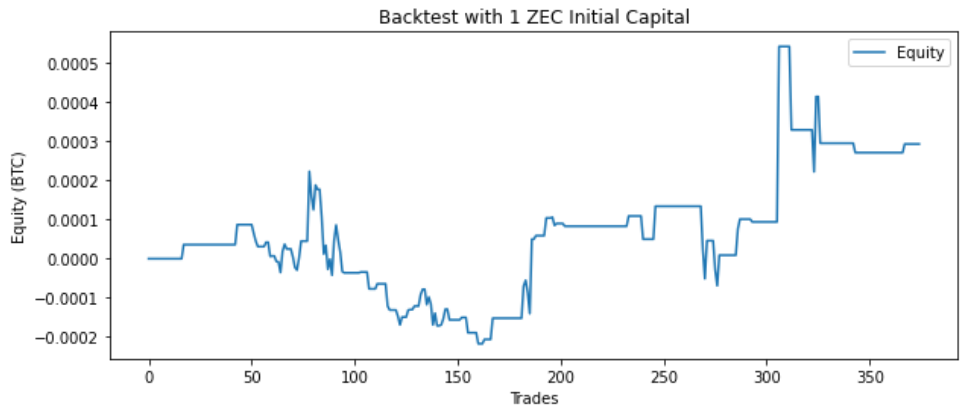

Figure 33. Performance of The XGB Model for ZEC-BTC in Augmented Test Data

| | |
|---|---|
| Testing Accuracy | 0.521277 |
| Net Profit | 0.0003430 |
| Number of Winning Trades | 21 |
| Number of Losing Trades | 22 |
| Total Days in Test | 66 |
| Percent of Profitable Trades | 48.84% |
| Avg Win Trade | 0.00004 |
| Avg Los Trade | -0.00002 |
| Largest Win Trade | 0.00015 |
| Largest Los Trade | -0.00010 |
| Profit Factor | 1.63 |

Table 7. Information Regarding k-NN Test on ZEC-BTC

| | |
|---|---|
| Testing Accuracy | 0.510638 |
| Net Profit | 0.0007550 |
| Number of Winning Trades | 87 |
| Number of Losing Trades | 85 |
| Total Days in Test | 66 |
| Percent of Profitable Trades | 50.58% |
| Avg Win Trade | 0.00004 |
| Avg Los Trade | -0.00004 |
| Largest Win Trade | 0.00020 |
| Largest Los Trade | -0.00014 |
| Profit Factor | 1.25 |

Table 8. Information Regarding RF Test on ZEC-BTC

| | |
|---|---|
| Testing Accuracy | 0.518617 |
| Net Profit | 0.000293 |
| Number of Winning Trades | 46 |
| Number of Losing Trades | 49 |
| Total Days in Test | 66 |
| Percent of Profitable Trades | 48.42% |
| Avg Win Trade | 0.00005 |
| Avg Los Trade | -0.00004 |
| Largest Win Trade | 0.00045 |
| Largest Los Trade | -0.00021 |
| Profit Factor | 1.14 |

Table 9. Information Regarding XGB Test on ZEC-BTC

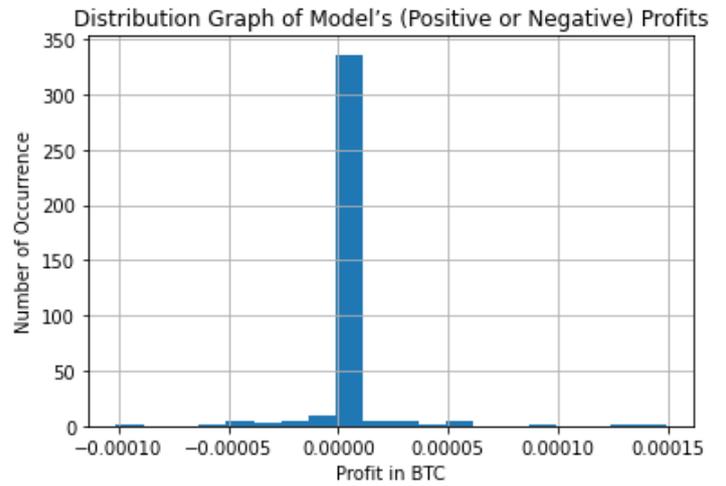

Figure 34. Distribution of Profits for k-NN on ZEC-BTC

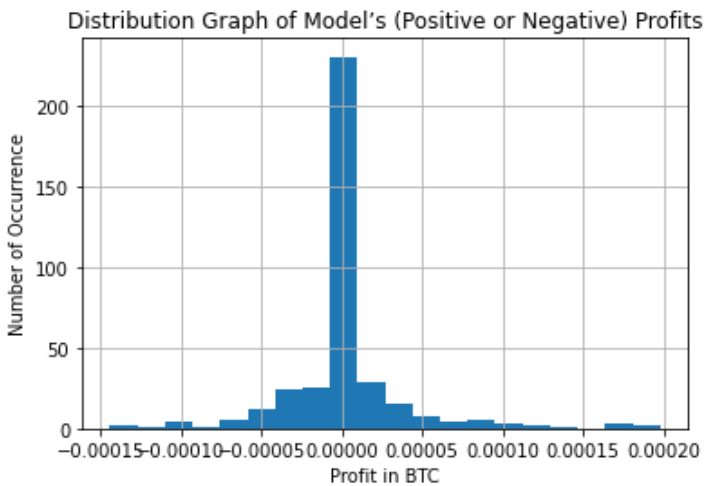

Figure 35. Distribution of Profits for RF on ZEC-BTC

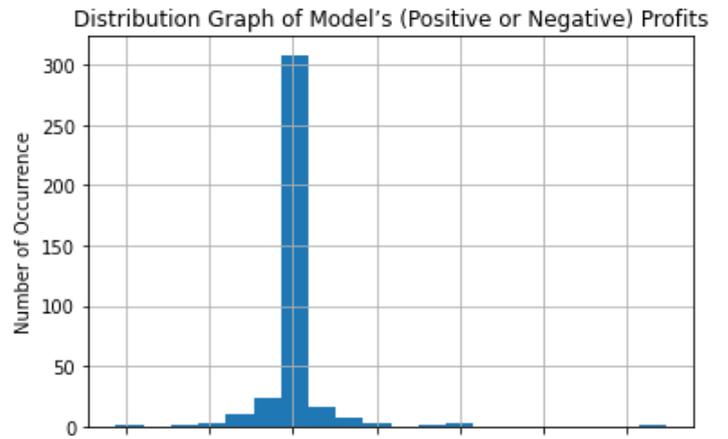

Figure 36. Distribution of Profits for XGB on ZEC-BTC

## VI. DISCUSSION

In this section we discuss how our models have been performing through different market conditions. We also talk about how these results challenge the Efficient Market Hypothesis (EMH) in the context of cryptocurrency markets and how one can implement practically these models and exploit the market. We also note the limitations and differences between response times of our models.

All our three studied cryptocurrency pairs show different market conditions through our models train phases (i.e. all of them have bullish, bearish and range movements), although frequency of these moves are not equal and this could make effects on the results. Let's compare test results on late April 2021 fall of ETH-USDT. By comparing figure 3 and figure 4 we see k-NN based strategy has not fallen in equity during that span, but figures 5 and 6 show that XGB and Random Forest have fallen there, although the overall smoothness of the last two models is better. Again, if we take figure 11 as a reference and compare figures 12 and 14 with it, we see how models have avoided bearish markets. Figure 14 shows a model which, although being positive in the final returns, does not look good for practical usage. Cumulative results in 66 days for all models show a positive return.

Based on the obtained results in this study it can easily be seen that augmenting the input data with technical indicators has a significant positive effect on the final returns for these models. The extent of this effect can be moving from a negative return to a positive one.

"The efficient market hypothesis (EMH) is a back-breaker for forecasters. In its crudest form it effectively says that the series we would very much like to forecast, the returns from speculative assets, are unforecastable. The EMH was proposed based on the overpowering logic that if returns were forecastable, many investors would use them to generate unlimited profits. The behaviour of market participants induce returns that obey the EMH, otherwise there would exist a 'money-machine' producing unlimited wealth, which cannot occur in a stable economy." (Timmermann & Granger, 2004) The point here is, what if some patterns exist in the market dynamics but are not visible to ordinary users and they could only be discovered through artificial intelligence based systems? If that would be the case, it's obvious that our formulation of EMH should change accordingly. Maybe the hidden patterns inside a market can be exploited to some extent and this extent is determined by the behaviour of the exploiters. But of course due to the laws of thermodynamics there will be a limit for this exploit. It can be definitely said due to the second law of thermodynamics, a system's entropy increases and this eventually makes that system unstable for exploitation. So, in this regard different algorithms can exist which compete with each other in different market dynamics and some of the times some of them can be superior to others and create profit. These algorithms efficiency can be related to their "intelligence" level in choosing the proper action in the market and also on how other participants in the market are behaving.

There were two important factors of tuning our studied models: Hyperparameters and Labelling Sensitivity Threshold. Each discussed model has its own hyperparameters and changing them affected the results significantly. Labelling Sensitivity Threshold was another consequential parameter in our experiments. This parameter defines to which extent the return should be, to be considered a positive return for the model. Usually, it should be at least greater than the exchange's fee to denote a profitable trade, but tweaking with it yields different results. One can use grid searching in available values for these tunings. There may be a need to reconfigure them from time to time.

As it has been shown in this project, models perform differently in different markets. Beside the generated profit, each machine learning model can have its own pros and cons. For example, in our experiments, k-NN usually took about 7.5 seconds to predict the label where Random Forest took about 0.13 seconds and XGB took only 0.008 seconds. These differences will make each of them preferable to another based on the contexts.

All of what has been discussed till now are theoretical arguments, implications of these models look very attractive but it definitely will bring up new issues and more research needs to be done. Many exchanges nowadays allow automatic traders to act in their provided markets. One can use these exchanges data and process them inside the introduced schemes and decides and trades based on them in hope of profit. As cryptocurrency markets are almost always available (i.e. 24/7) using a dedicated server can find trade opportunities and acts on them automatically.

### VII. CONCLUSIONS AND FUTURE WORKS

The impact of artificial intelligence's applications in many areas are promising for a more efficient and prosperous future. In this study we looked at three different machine learning approaches to help investors to make their decisions in some new emerging international markets in a more data driven and autonomous manner. We showed how using technical indicators can improve the results of the prediction system. We also addressed a simple strategy design framework to use these models. Although all of our models showed positive returns and a maximum of 1.60 profit factor for ZEC-BTC by k-NN in 66 days and a minimum profit factor of 1.12 for LTC-BTC by Random Forest, it's obvious more research needs to be done in this area. The resulting strategies still lack "smoothness" in their equity graphs and hence showing large potential risks to be implemented. Designing a full autonomous trading system surely involves more concerns than the ones we had simplified in this research work, like market liquidity issues. We also discussed how these new and uprising technological advancements can cast a shadow on some long lasting hypothesis in finance like Efficient Market Hypothesis.

As we can see, there seems a predictability potential in a highly complex system like financial markets by means of machine learning. For future works, our suggestions include:

1. Combining Fundamental Information with Technicals to improve the accuracy
2. Ensembling different approaches in machine learning to decrease the bias of the whole system
3. Using social networks data streams to obtain an accumulated view on public opinion on different assets
4. Using Deep neural networks to feature extraction from raw data
5. Using Deep Reinforcement Learning to design sophisticated strategies directly with the aim of enhancing performance
6. Using machine learning approaches for risk management in a collateral system to decision making

Besides what we have discussed about financial markets, it seems machine learning models can be used in many other chaotic natured problems which share some of their data characteristics with financial data. These fields could include supply chain support, Business affairs with public opinions, public views on political issues and many other use cases.

## Appendix A: Used technical indicators and their formulas

In this appendix we introduce the technical indicators used in this project and their respective formulas.

Commodity Channel Index (CCI):

$$CCI = \frac{Typical\ Price - MA}{0.015 * Mean\ Deviation}$$

where:

$$Typical\ Price = \sum_{i=1}^{P}((High + Low + Close)/3)$$
$$P = Number\ of\ Periods$$
$$MA = Moving\ Average$$
$$Moving\ Average = (\sum_{i=1}^{P} Typical\ Price)/P$$
$$Mean\ Deviation = (\sum_{i=1}^{P} |Typical\ Price - MA|)/P$$

We have used this indicator in 14 and 30 periods in this project.

Relative Strength Index (RSI):

$$RSI_{Step\ One} = 100 - \left[\frac{100}{1 + \frac{Average\ Gain}{Average\ Loss}}\right]$$

The average gain or loss used in the calculation is the average percentage gain or loss during a look-back period. The formula uses a positive value for the average loss.

Once there is first step data available, the second part of the RSI formula can be calculated. The second step of the calculation smooths the results:

$$RSI_{Step\ Two} = 100 - \left[\frac{100}{1 + \frac{(Previous\ Average\ Gain * (Period - 1)) + Current\ Gain}{-((Previous\ Average\ Loss * (Period - 1)) + Current\ Loss)}}\right]$$

We have used this indicator in 14 and 30 periods in this project.

Directional Movement Index (DMI):

$$DX = \left(\frac{|DI^+ - DI^-|}{|DI^+ + DI^-|}\right) * 100$$

where:

$$DI^+ = \left(\frac{Smoothed\ (DM^+)}{ATR}\right) * 100$$
$$DI^- = \left(\frac{Smoothed\ (DM^-)}{ATR}\right) * 100$$
$$DM^+ (Directional\ Movement) = Current\ High - Previous\ High$$
$$DM^- (Directional\ Movement) = Previous\ Low - Current\ Low$$
$$ATR = Average\ True\ Range$$
$$Smoothed\ (x) = \sum_{t=1}^{Period} x - \frac{\sum_{t=1}^{Period} x}{Period} + CDM$$
$$CDM = Current\ DM$$

We have used this indicator with period=14 in this project.

Moving Average Convergence Divergence (MACD):

$$MACD = EMA_{12\ Period} - EMA_{26\ Period}$$

Bollinger Band®:

$$Boll = \frac{Boll_U + Boll_D}{2}$$
$$Boll_U = MA(TP, n) + m * \sigma[TP, n]$$
$$Boll_D = MA(TP, n) - m * \sigma[TP, n]$$

where:

$$Boll_U = Upper\ Bollinger\ Band$$
$$Boll_D = Lower\ Bollinger\ Band$$
$$MA = Moving\ Average$$
$$TP\ (Typical\ Price) = (High + Low + Close)/3$$
$$n = Number\ of\ Days\ in\ Smoothing\ Period\ (Typically\ 20)$$
$$m = Number\ of\ Standard\ Deviations\ (Typically\ 2)$$
$$\sigma[TP, n] = Standard\ Deviation\ over\ Last\ n\ Periods\ of\ TP$$